\documentclass[conference]{IEEEtran}
\IEEEoverridecommandlockouts
\usepackage{cite}
\usepackage{amsmath,amssymb,amsfonts}
\usepackage{algorithmic}
\usepackage{graphicx}
\usepackage{textcomp}
\usepackage{xcolor}

\usepackage{multirow}

\def\BibTeX{{\rm B\kern-.05em{\sc i\kern-.025em b}\kern-.08em
    T\kern-.1667em\lower.7ex\hbox{E}\kern-.125emX}}
\begin{document}

\title{Gathering GitHub OSS Requirements from Q\&A Community: an Empirical Study \\
\thanks{*Corresponding author: xjmao@nudt.edu.cn }
\thanks{ This work is sponsored by the National Key Research and Development Program of China under Grant 2018YFB1004202.}
}

\author{\IEEEauthorblockN{Hao Huang, Yao Lu, Xinjun Mao\IEEEauthorrefmark{1}} \IEEEauthorblockA{Key Laboratory of Software Engineering for Complex Systems\\ College of Computer, National University of Defense Technology, Changsha 410073, China \\ Email: dongfangyier163@163.com, \{luyao08, xjmao\}@nudt.edu.cn}} 

\maketitle

\begin{abstract}
Cross-community collaboration can exploit the expertise and knowledges of crowds in different communities. Recently increasing users in open source software (OSS) community like GitHub attempt to gather software requirements from question and answer (Q\&A) communities such as Stack Overflow (SO). In order to investigate this emerging cross-community collaboration phenomenon, the paper presents an exploratory study on cross-community requirements gathering of OSS projects in GitHub. We manually sample 3266 practice cases and quantitatively analyze the popularity of the phenomenon, the characteristics of the gathered requirements, and collaboration behaviors of cross-community. Some important findings are obtained: more than half of the requirements gathered from SO are enhancements and the majority of the gathered requirements are non-functional requirements. In addition, OSS developers can directly obtain related solutions and contributions of the gathered requirements from SO in the gathering process.
\end{abstract}


\section{Introduction}
GitHub is a social collaborative software development community \cite{b1}, in which developers over Internet work together to develop open source software (OSS) in term of social and collaborative development activities like fork, pull requests and issue, etc. \cite{b2}. Users in GitHub typically provide their contributions in term of submitting their codes, issues, and comments, etc., among of which issues can take diverse forms, like bugs, enhancements, features, etc., and represent specific kind of requirements for open source software. Based on the issues and discussions of OSS projects, developers can contribute their codes to solve the issues. Obviously, GitHub provides flexible and effective mechanisms to gather software requirements from developers in GitHub, which is significant for promoting the collaborative development and long-term evolution of open source software. For example, in the past five years, more than 2400 developers of the GitHub project \textit{TensorFlow} have contributed 24103 issues which result in 106 version evolution. 

Stack Overflow (SO) is one of the most popular question and answer (Q\&A) site, in which users around the world discuss the coding issues, share the code examples and exchange software development knowledges and expertise \cite{b3}. Recently users in SO and GitHub overlap, which means that many users in GitHub are also the users of SO, and vice versa. Silvestri et.al. \cite{b4} explored account association between Stack Overflow and GitHub, and matched more than 600,000 users between the two platforms. Such trend implies that users in SO and GitHub may have the same OSS projects’ background and development expertise. Lee et. al. \cite{b5} analyzed developer interests across multiple social collaborative platforms, and found 39\% developers share common interests in SO and GitHub.

Obviously, SO and GitHub are two communities with different design objectives. However, developers in these communities actually can collaborate with each other to solve specific development problems such as requirement gathering, code reusing. For example, developers in GitHub can reuse the code snippets present by users in SO to develop open source software. The research \cite{b6} shows that 69 vulnerable code snippets in SO are reused in 2859 GitHub projects. Such collaboration may cause the secure risks for open source software in GitHub \cite{b7}. Some researches determine the potential defects of the given source codes in term of scoring the code defects in SO and GitHub, and analyzing their correlations \cite{b8}.

Software requirements describe the expectations of users towards the software product. Requirements Engineering (RE), i.e., requirements elicitation, analysis, specification, validation, and management, is a fundamental phase of every software development project \cite{b9}. Requirements elicitation is also called requirements gathering, which is a development activity to collect and define the requirements for specific software system or project. In open source software community like GitHub, software requirements of OSS are represented as issues that may take multiple forms, such as bugs, enhancements, new feature suggestions, etc. Typically, they are presented by the community crowds that participate in the development of the OSS projects. 

\begin{figure*}[htbp]
	\centerline{\includegraphics[width=6.7in]{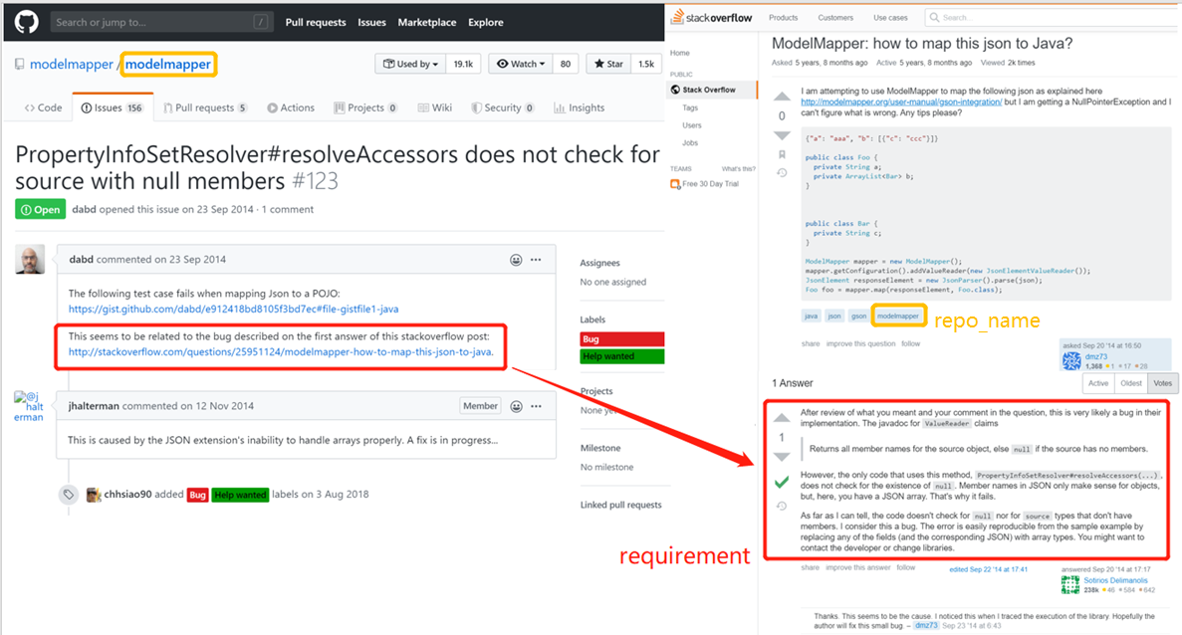}}
	\caption{An example of gathering software requirements of the GitHub project modelmapper/modelmapper from SO.}
	\label{fig01}
\end{figure*}

Recently, more and more evidences indicate an emerging phenomenon that the developers in GitHub gather the OSS projects requirements from the shared knowledges published in SO. Users in SO provide their contributions on software requirements of OSS projects in GitHub by discussing their confusion and difficulties encountered when using OSS projects, and the functions they want to enhance or add when applying OSS projects. Developers in GitHub can exploit the cross-community shared knowledges to discover these potential and valuable software requirements, with which they fix bugs and enrich the functionalities of OSS projects. Such kind of cross-community software development practices and collaboration enable OSS project developers in GitHub to gather more valuable software requirements from the crowds outsides of the GitHub. For example, according to our analysis, 1892 issues of GitHub project TensorFlow contains Stack Overflow’s links, many of them are software requirements gathered from the discussions in Stack Overflow.

Fig.~\ref{fig01} depicts an example of gathering requirements of the \textit{ModelMapper} OSS project across communities. User \textit{Dabd} in GitHub found a \textit{ModelMappe’s} bug presented by user \textit{dmz73} and \textit{Sotirios Delimanolis} in SO. \textit{Dabd} opened an issue in GitHub, briefly introduced the content of bug and the source of the question (SO answer), and referenced the post's link in the issue. 

According to our observation, gathering requirements of OSS projects across communities is increasing popular in recent years, and plays an important role to promote the development of OSS projects in GitHub. More and more OSS projects in GitHub gather their software requirements in term of referring to the requirement knowledges and suggestions of users in SO. Many developers in GitHub browse the posts in SO, and create a new issue in the GitHub repository when they find interesting knowledges related with their OSS projects. These development behaviors have gradually become an important way to gather software requirements of OSS projects in GitHub. However, we still lack of the in-depth understanding of this emerging phenomenon and the intrinsic development practices and social collaborations between the users across the two communities. We also know little about the characteristics of the gathered software requirements in term of cross-community collaboration. 

The remaining sections are organized as follows. Section 2 introduces the related work. Section 3 presents the research questions and details the study method. Section 4 presents our empirical analysis results and introduces some important findings. Section 5 discusses the extended implications and threats to validity. Finally, Section 6 concludes the paper and summarizes our contributions.

\section{Related Work}
This section introduces the related work on the software requirements gathering in GitHub and the cross-community software development collaboration.
\subsection{Software Requirement Gathering and Management in GitHub}
GitHub provides a flexible issue tracking system to help users to gather and process software requirements for OSS projects \cite{b10}\cite{b11}. As mentioned from Liao et al. \cite{b12}, the time distribution of issue commits followed a three-period development model, and the "Feature Request" was the tag with the highest frequency (56\%) in the first period, so this stage mainly corresponds to the process of obtaining requirements in the project life cycle. Our work is focused on analyzing the process of obtaining requirements in GitHub issues.

GitHub projects have a readme file to describe its main functionalities, it is an interesting information source for gathering requirements. Roxana et al. \cite{b13}\cite{b14} used GitHub readme files as an information source for requirements elicitation, proposed a method to provide knowledge to requirements engineers from a viewpoint of domain. On the basis of the above methods, they performed a just-in-time requirement elicitation by analyzing the text of the issue in similar projects, presented a tool \cite{b15} to help eliciting requirements related information and further extract the software requirement pattern \cite{b16} of OSS projects. They used the issue perspective to model more details about features (e.g. bugs or enhancements) which were refined by using its comment perspective. But their data collection and procession were based on the ReadMe files, so it is easy to be limited by the quality and completeness of the readme files. In addition, readme files are generally not updated frequently, which will cause the outdated requirements to be gathered. This paper is directly from the issue perspective and explains the practice of gathering requirements across communities.

\subsection{Cross-Community Development Collaboration}
There are increasing interests on the study of cross-community social collaboration phenomenon and their contributions to OSS projects. Many cross-community researches start with the comparative study of user interaction networks, investigate the correlations between users and their interactions \cite{b4}, and analyze the user’s behaviors through the associated account. For example, the researches \cite{b17}\cite{b18} mined the heterogeneous information both Q\&A sites and OSS communities to model and evaluate the programming ability and expertise of developers across communities, and estimated the competency of developers in answering the questions posted in SO. Vasilescu et al. \cite{b19} investigated the interplay between asking and answering questions on Stack Overflow and committing changes to open source GitHub repositories. They found that active GitHub committers asked fewer questions, but provided more answers than others, and that SO activity rates are related to code change activities in GitHub. Xiong et al. \cite{b20} found similar phenomenon, pointed out that the active issue committers in GitHub were also active question askers in Stack Overflow. For most of developers, the topics of their contents in GitHub were similar to that of their questions and answers in Stack Overflow and developers’ concerns in Stack Overflow shifted over the time of their current participating projects in GitHub. The previous researches are not focused on the practice of gathering requirements. Our work fills this gap to get some insights about the cross-community requirements gathering between GitHub and SO.

\section{METHODOLOGY}
In this section, we first give our research questions, and then we present our research datasets and methods to support the research on the designed questions.
\subsection{Research Questions}
The goal of our study is to explore the cross-community requirement gathering phenomenon and the insights of such collaborative development practices. In order to achieve the goal and guide the study, we design three research questions to elaborate on our research focuses and decompose the whole research work. 

\textbf{RQ1: What’s the popularity of cross-community software requirements gathering? }
This research question intends to investigate the current practices of cross-community software requirements gathering, and gain the intuitive understanding of such emerging phenomenon. We further decompose the question as the following fine-grained sub-questions:

RQ1.1: What’s the popularity of cross-community linking behavior?

RQ1.2: What’s the popularity of gathering software requirements from SO in OSS projects in GitHub? 

\textbf{RQ2: What are the characteristics of OSS requirements gathered from SO?}
This research question aims to investigate the particularity of the OSS requirements gathered from SO. It is helpful to discriminate the software requirements that are more suitable for cross-community gathering. 

\textbf{RQ3: How do users gather cross-community software requirements?}
This research question attempts to examine the insights of the requirement gathering from the viewpoint of user contributions and cross-community behaviors, including the contributions of different type of participants in GitHub and SO to gather cross-community requirements, and the user’s specific behaviors of gathering requirements in two communities. Sub-questions: 

RQ3.1: How much do different users in SO and GitHub contribute to gathering cross-community needs?

RQ3.2: What specific behaviors do users in SO and GitHub use to gather cross-community requirement?

\subsection{Research Datasets}
We use two open datasets from GitHub and SO, i.e., GHTorrent \cite{b2}\cite{b21} and SOTorrent \cite{b22}. The GHTorrent is a mirror and index data from the GitHub API \cite{b2}. It provides diverse data about OSS projects, development activities and collaborations, e.g., pull requests, commits, etc. The SOTorrent is an open dataset based on the official SO data dump \cite{b22}. We used the SOTorrent datasets that contain all data before December 2019. 

Based on the GHTorrent that contains 1790 repositories, 327,313 commits, 622,323 issues, 17,371 Wikis citing SO links in GitHub, we finally collect 3266 issues in total for manually annotation, 1034 issues of them are software requirements. Moreover, we collect 939 effective SO post data in SOTorrent based on the SO link information in 1034 issues in GHTorrent. In the following, we introduce the data processing methods for supporting the analysis. 
\subsection{Research Methods}
\subsubsection{Collecting and Pre-processing GitHub Issue Datasets}
\begin{figure*}[htbp]
	\centerline{\includegraphics[width=6in]{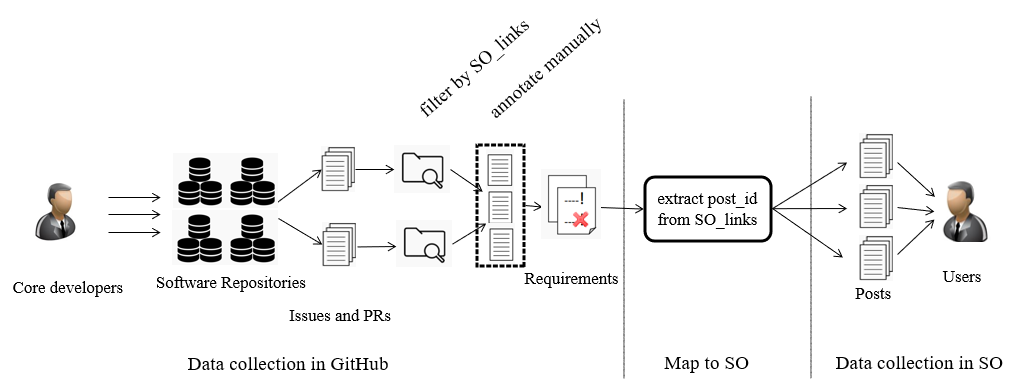}}
	\caption{Methods of data collection and processing.}
	\label{fig02}
\end{figure*}
Data collections comprise of two parts: collecting GitHub data and collecting SO data (see Fig.~\ref{fig02}). We use GitHub API v3 to retrieve issue data. The quantity of search results limited to 1000 so we use stratified sampling to select samples randomly every six months and obtain 3266 issues as raw data in total for manually annotation. Preprocessing on raw dataset are necessary in order to support the follow-up study activities, e.g., filtering out non-English issues, non-requirements issues and the issues in which SO links are contained in comment part. We obtain the required issue data that describe the information of the cross-community requirement gathering from the GitHub repository and explicitly contain SO links describing the associations between the issue in GitHub and the Q\&A in Stack Overflow.

\subsubsection{Annotating GitHub Data}
A lot of semantics information of issue data that are required for analysis are missing or unclearly described in raw datasets. For example, whether the issue describes a functional requirement or non-functional requirement. We need to read and understand the issue data, get the expected semantics information, annotate and add these information manually in issues datasets. If the issue is used to feedback repositories’ bugs, defects, enhancements, new feature suggestions, and other software development related issues, we will determine this issue is used to gather requirements and record other attributes of requirements. For some issues with brief content and few text descriptions, we can further judge by visiting the issue URL and referring to the comments and context of the issue. We can also click on the SO links citing in the issue, jump to SO website to browser the corresponding post contents. Finally, we obtain 1034 software requirements with annotation for subsequent data analysis.

\subsubsection{Intersecting issue-post Link Pairs}
In order to analyze the requirement associations and development collaborations across communities, we need to correlate the issue data from GitHub with the Q\&A dataset in SO. We extract the complete SO links from the issue data, obtain the post\_id information from the links as input to query the SOTorrent dataset in order to acquire the post’s information such as question, answers, comments, post creation time, etc. associated with the issue.
\subsubsection{Analyzing Associated Data}
We analyze the above data in GitHub and SO and get some meaningful insights about the designed three research questions. The details show in next section.

\section{Findings}
This section introduces the analysis based on the collected datasets of GitHub and SO, and details some important insights and findings. 

\subsection{RQ1: The popularity of cross-community software requirements gathering}
Gathering software requirements by cross-community development collaboration is a new way of eliciting software requirements for OSS projects in GitHub. Such phenomena and practices have gained great attentions of software developers in GitHub and arise in increasing number of OSS projects. We investigate the popularity of the linking behavior between GitHub and SO, and the popularity of gathering software requirements from SO in OSS projects in GitHub to assess the popularity of this practice.
 
\begin{figure}[htbp]
	\centerline{\includegraphics[width=3in]{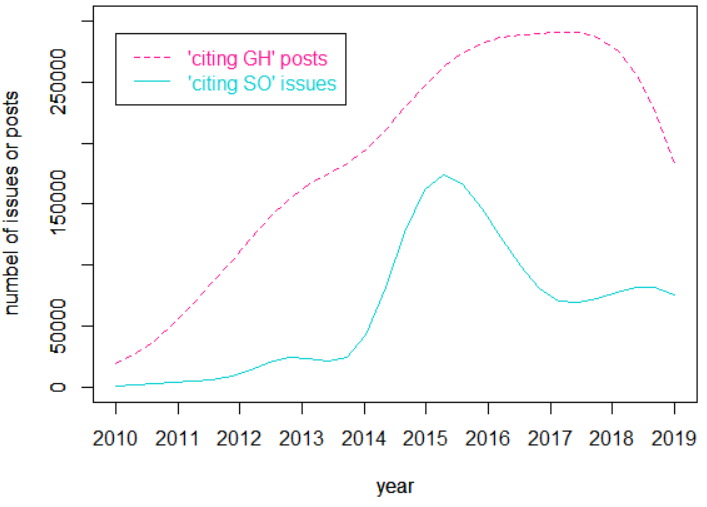}}
	\caption{The number of ”citing GH” posts and ”citing SO” issues posted each year.}
	\label{fig03}
\end{figure}

\begin{figure}[htbp]
	\centerline{\includegraphics[width=3in]{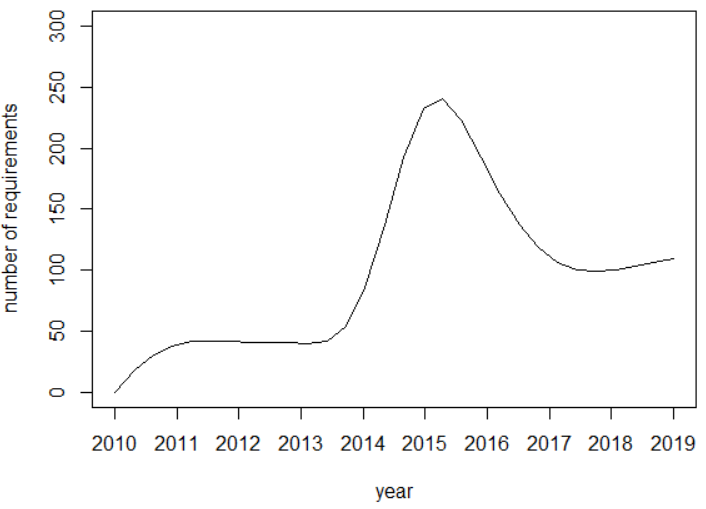}}
	\caption{The number of requirements gathered from SO posted each year.}
	\label{fig04}
\end{figure}

We calculate the frequency of GitHub issues with SO links and SO posts with GitHub links. Fig.~\ref{fig03} shows the number of ‘citing GH’ posts and ‘citing SO’ issues every year in two communities. Overall, SO has more cross-community linking behaviors than GitHub, because SO is a Q\&A community and cites more external questions. GitHub is a software development platform, cross-community linking behavior is relative less. From the Fig.~\ref{fig03} we can see that the cross-community linking behavior increased rapidly in 2015. SO's cross-community linking behavior gradually slowed down in 2016 and reached a steady growth in 2017, while GitHub’s cross-community linking behavior has been slowing in recent years. However, this cross-community linking behavior is very common, with tens of thousands of new links appearing between GitHub and SO every year.
Then we analyze the frequency of 1034 gathering requirements across communities every year. Fig.~\ref{fig04} shows the number of requirements gathered from SO every year. We can observe that the practice of gathering software requirements appeared in 2011, it grows the fastest in 2015, and has gradually reached a steady growth in recent years.
In addition, we investigate how popular are OSS projects gathering software requirements from SO. We analyze data related to the OSS projects to which the issue belongs, and found that 864 different open source projects were involved in 1034 issues, which shows that this practice is very common, many repositories have gathered requirements from SO. For example, Flutter is an OSS project that has been active in recent years which has 37k issues. We find that 1425 issues in Flutter gathered requirements from SO.

\noindent
\fbox{%
	\parbox{3.35in}{%
					\textit{Finding 1: }The practice of gathering software requirements across communities is prevalent. Many OSS projects have gathered software requirements from SO.
	}%
}

\subsection{RQ2: The characteristic of the OSS requirements by cross-community gathering}
We analyze the characteristics of issues with cross-community requirements and issues without cross-community requirements. We investigate which types of software requirements are more likely to be gathered in term of cross-community collaboration, and whether these software requirements are more focused on functional or non-functional requirements. 

In this investigation, we mainly focus on the: classification of requirements according to issue labels (e.g. bug, feature, enhancement.), distribution of functional and non-functional requirements. The specific analysis is as follows:

\subsubsection{Bug, Feature or Enhancement?}
\begin{figure}[htbp]
	\centerline{\includegraphics[width=3.3in]{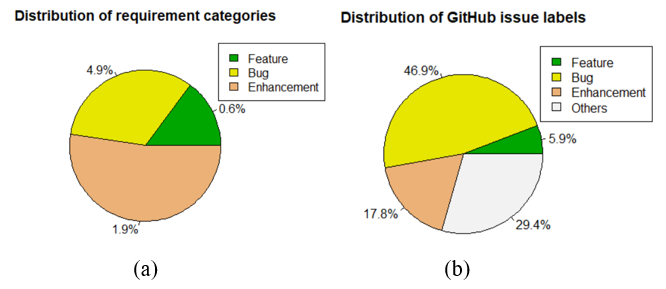}}
	\caption{The distribution of requirements categories by issue label.}
	\label{fig05}
\end{figure}

In general, labeling issues is more conducive to issues’ understanding and management. Issue tracking system in GitHub can define a label list to classify issues, facilitate issue management, and favor the resolution of issues \cite{b23}. By analyzing the issue's label list, we find that the most frequently used labels are: enhancement, bug, feature. The enhancement label represents the enhancement of the original function, and the feature represents the addition of a new function. In addition to these labels, 36 issues are labeled with other labels such as help wanted, component and so on, 119 issues have no labels. We manually classify this part of the data according to enhancement, bug, and feature. Issues outside these three categories will not be discussed. The final distribution of requirements categories is shown in the Fig.~\ref{fig05} (a) below. 
The previous research \cite{b11} shows that the most used tags in GitHub issues are bug (46.9\%), enhancement (17.8\%), and feature (5.9\%), as is shown in Fig.~\ref{fig05} (b). But in our research, we find that among cross-community requirements, only about 32.67\% of the requirements corresponds to bug, while 52.42\% corresponds to enhancement, and 14.92\% corresponds to feature. Thus, we find that more and more enhancement requirements have been gathered from SO. We consider that due to the increasing discussion in SO, it has gradually become a huge knowledge base derived from the Internet crowds. More people directly find some similar questions by searching when they encounter problems, instead of asking questions directly, so more and more posts are cited in GitHub for gathering enhancement requirements.
\subsubsection{Functional Requirement or Non-Functional Requirement?}
We classify requirements into Functional Requirement (FR) and Non-Functional Requirements (NFR), which are the basic classification of software requirements. FR define a function that a system or system element must be qualified to perform and describes the behavior of the system as it correlates to the system’s functionality. And NFR is restricted to a set of specific qualities other than functionality: such as usability, reliability, maintainability, extensibility, scalability and security. For example, a question: “What information must be stored in the database.” can be categorized as a FR. “The memory used by the software does not exceed 10Mb.” can be manually classified as an NFR.

The description about FR and NFR is mentioned in reference \cite{b24}\cite{b25}. According to reference \cite{b24}, we think that Feature belongs to FR, and we more agree with reference \cite{b25} to believe Bug belongs to the reliability of NFR. We only need to annotate the Enhancement to determine its type.

\begin{table}[htbp]
	\centering
	\caption{THE NUMBER OF DIFFERENT TYPE OF REQUIREMENTS}
	\begin{tabular}{llllll}
		\hline
		\#type & Enhancement & Bug &Feature  &\#sum &\#rate  \\ \hline
		FR & 118 & 0 & 97 & 215 & 27\%  \\ 
		NFR & 373 & 208 & 0 & 581 & 73\% \\ \hline
	\end{tabular}
	\label{tab01}
\end{table}
As shown in Table~\ref{tab01}, we found that most of the requirements discussed in SO are non-functional requirements (73\%), and there are fewer functional requirements (27\%). These non-functional requirements include maintainability, efficiency, availability, reliability, etc., we find most of NFRs gathered from SO are questions about system safety and performance testing. OSS developers in GitHub mainly focus on functional requirements, while non-functional requirements are difficult to gather. Through the Q\&A community, it is easier to gather non-functional requirements from the crowds, which is of great help to improve the security, reliability, and stability of the software system. The developers can not clearly express these non-functional requirements, so they ask questions in SO. After full discussion, some developers move this question to the corresponding GitHub software repository.
\noindent
\fbox{%
	\parbox{3.35in}{%
		\textit{Finding 2:} More than half software requirements gathered from SO is enhancement requirements, followed by bug and feature. We annotate requirements from the dimensions of FR and NFR, and find that most cross-community requirements are non-functional requirements about security and testing.
	}%
}

\subsection{RQ3: The user behaviors of cross-community software requirements gathering}
In this research question, we further investigate how OSS developers gather software requirements across SO and GitHub. We analyze this question from the perspective of participants’ contributions and specific behaviors of gathering cross-community requirements.

\subsubsection{Participants’ contribution of gathering cross-community requirements in SO and GitHub}
We categorize users in SO and GitHub, and investigate different type of participants’ contribution of gathering cross-community requirements. In SO, we classify users according to reputation values and privileges of user. In GitHub, we classify users according to the association between the issue author and the repository which the issue belongs to. The following are the details and results:

\textbf{Users’ contribution of gathering cross-community requirements in SO: }Reputation is a rough measurement of how much the community trusts you. High-reputation users contribute more to community development. In SO, the primary way to gain reputation is by posting good questions and useful answers. It represents the higher reputation users have, the more discussions users participate in, the more active users are, and the more experience and knowledge they gain. We classify SO users according to privileges and get the following four categories: new user, active user, established user and trusted user. The description of privileges and the corresponding user's reputation value are shown in Table~\ref{tab02}.
\begin{table}[htbp]
	\centering
	\caption{User type and DESCRIPTION in SO}
	\begin{tabular}{p{0.6in}p{0.4in}p{2in}}
		\hline
		\#User Type & Reputation  & Privilege  \\ \hline
		trusted user & 20k+ &Expanded editing, deleting and undeleting privileges.\\ 
		established user & 1k-20k &Create tags and edit questions and answers.\\ 
		active user & 10-1k&Post more links, answers protected questions, create gallery chat rooms.\\
		new user & 1-10 &Create posts.\\ \hline
	\end{tabular}
	\label{tab02}
\end{table}
\begin{figure}[htbp]
	\centerline{\includegraphics[width=3in]{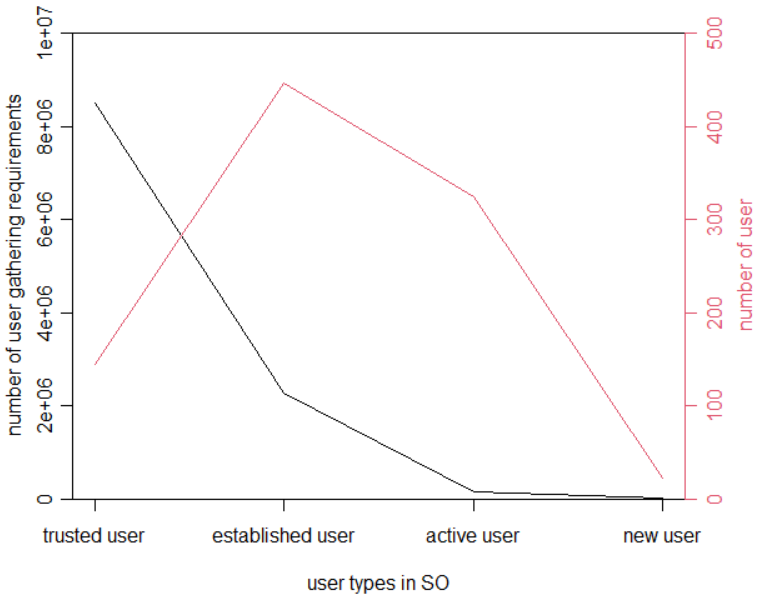}}
	\caption{The number of various types of users in SO and their contributions of gathering cross-community requirements.}
	\label{fig06}
\end{figure}

We calculate the total number of various types of users in SO and the various types of users’ contributions of gathering cross-community requirements. As shown in the Fig.~\ref{fig05}, trusted user in SO contribute the most to gathering cross-community requirements although the amount is the least. The contribution of established user and active user in gathering cross-community requirements is proportional to their respective numbers. Obviously, new users are not active whether in gathering cross-community requirements or participating in other discussions on SO. This shows that the new practice of gathering requirements across communities is mainly driven by active users.

\textbf{Users’ contribution of gathering cross-community requirements in GitHub:} The relationship between issue participants and repository is called author association in GitHub. There are five associations defined in GitHub: OWNER, MEMBER, CONTRIBUTOR, COLLABORATOR, NONE. Table~\ref{tab03} shows the detailed description of author association. According to the description of association, we classify both OWNER and MEMBER as OWNER, NONE as USER, and investigate the number of cross-community requirements gathered by different types of users in GitHub.

\begin{table}[htbp]
	\centering
	\caption{AUTHOR AND REPOSITORY ASSOCIATION DESCRIPTION}
	\begin{tabular}{p{0.8in}p{2.2in}}
		\hline
		\#author\_association & description    \\ \hline
		OWNER & The owner of the repository.\\ 
		COLLABORATOR & Author has been invited to collaborate on the repository. \\ 
		MEMBER & The member of the organization that owns the repository. \\ 
		CONTRIBUTOR & Author has previously committed to the repository. \\ 
		NONE & Author has no association with the repository. \\ \hline
	\end{tabular}
	\label{tab03}
\end{table}

\begin{figure}[htbp]
	\centerline{\includegraphics[width=2.5in]{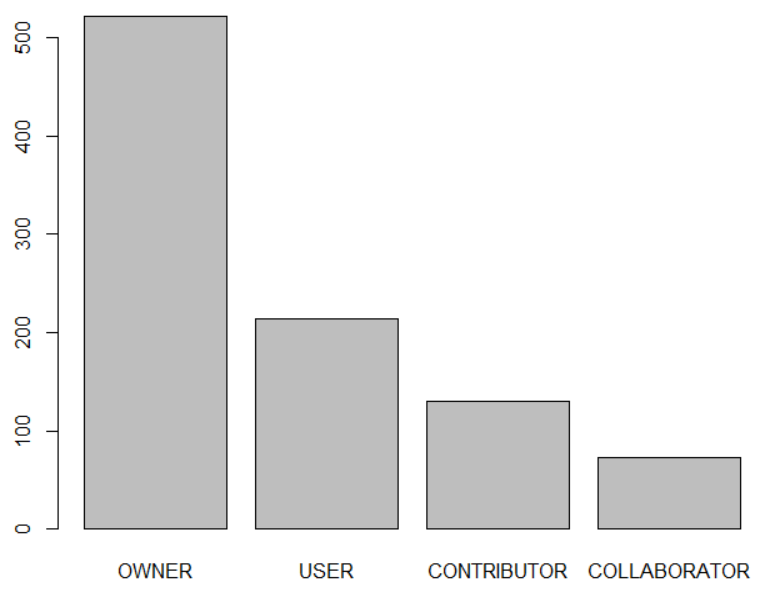}}
	\caption{The number of cross-community requirements gathered by different types of users in GitHub.}
	\label{fig07}
\end{figure}
As shown in Fig.~\ref{fig07}, most cross-communities requirements gathered by OWNER (55.5\%), it shows that repository owners are more concerned about their own OSS repositories. Moreover, USER has no association with the repository, but there are more contributions (22.8\%) of gathering cross-community requirements than collaborators (7.8\%) and contributors (13.9\%).

This shows that the new practice of gathering requirements across communities is mainly driven by users with higher permission in GitHub repository. However, the contributions made by other users who are not associated with the repository cannot be ignored.

\noindent
\fbox{%
	\parbox{3.35in}{%
			\textit{Finding 3: }Trusted users in SO with high reputation contribute the most to cross-community requirements gathering. Active repository owners in GitHub pay more attention to gathering cross-community requirements.
	}%
}

\subsubsection{Specific behavior of gathering cross-community requirements in SO and GitHub}
\begin{figure}[htbp]
	\centerline{\includegraphics[width=3.3in]{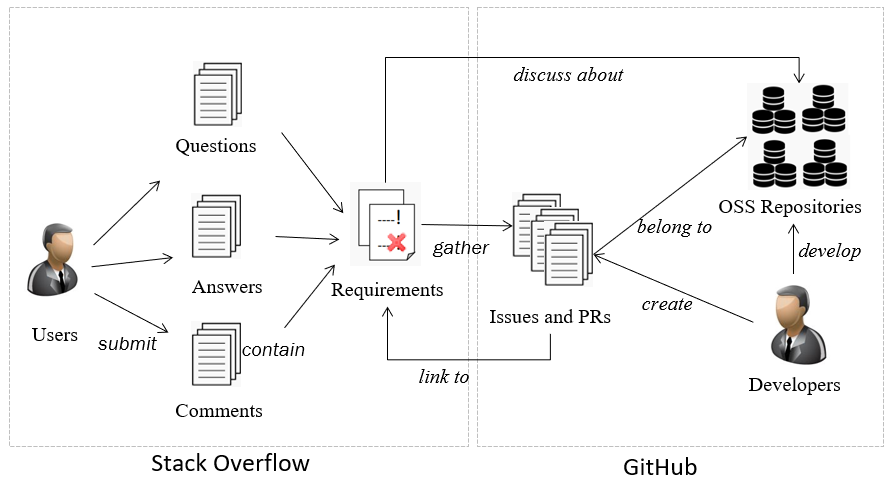}}
	\caption{Development collaborations to gather software requirements across GitHub and Stack Overflow.}
	\label{fig08}
\end{figure}

Fig.~\ref{fig08} illustrates the development collaborations of gathering OSS requirements across GitHub and SO. The main cross-community behavior of users is about gathering requirements from SO to GitHub. We can dispose this behavior into two parts, one is that users in SO discuss requirements about GitHub OSS projects, and the other is that users create new issues in GitHub and link to SO posts. We investigate the user behavior of gathering requirements across SO and GitHub, and find two specific behaviors: Referring GitHub OSS repository's name in SO posts, such as using repository's name as a post's tag; Citing multiple SO links in a GitHub issue of gathering cross-community requirements.

We define a new statistical measurement called Response time of gathering requirements, it represent the time period from the creation time of SO post of discussing requirements to the creation time of GitHub issue. To clearly illustrate our results, we used box plots in our research. In the box plot, five lines from top to bottom indicate the maximum value, the upper quartile (75\%), the median of the sample, the low quartile (25\%) and the minimum value. All data points above the top line or below the bottom line are outliers.

\textbf{Referring GitHub OSS repository's name in SO posts of discussing cross-community requirements:} In Fig.~\ref{fig01} of section 1, the SO post's author (\textit{dmz73}) used the GitHub repository name (\textit{ModelMapper}) as a tag of the question, which may help OSS developers (\textit{Dabd}) to find this requirement faster. We explore how many SO posts cited in GitHub also refer GitHub repositories and whether such behavior shorten the response time of gathering cross-community requirements. We find that in all 939 issue-post associated data, there are 251 SO posts referring GitHub repository's name, 52 of which refer in title, 109 in body, and 90 posts use repository's name as tags.

\begin{figure}[htbp]
	\centerline{\includegraphics[width=3.3in]{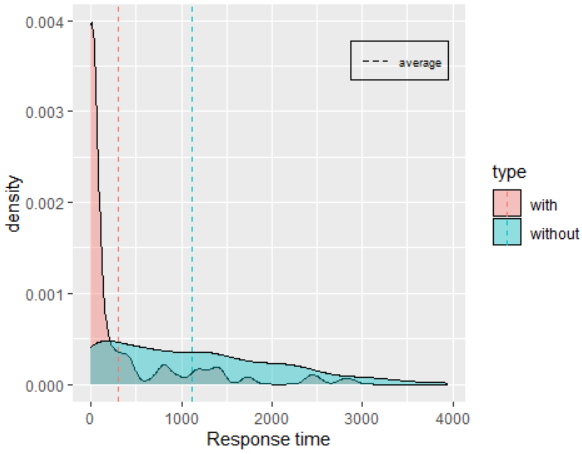}}
	\caption{Distribution of response time of gathering requirements from SO posts with or without GitHub repository’s name.}
	\label{fig09}
\end{figure}
\begin{figure}[htbp]
	\centerline{\includegraphics[width=2.5in]{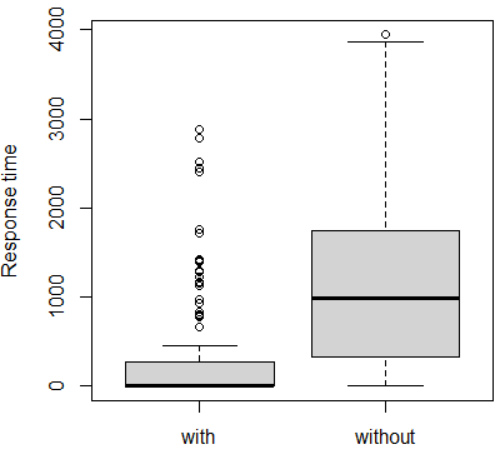}}
	\caption{Response time of gathering requirements from SO posts with or without GitHub repository’s name.}
	\label{fig010}
\end{figure}
We analyze response time of cross-community requirements with or without repository's name to answer this question. Calculate the distribution of response time of gathering requirements referring the repository's name.

As shown in Fig.~\ref{fig09} the distribution of response time of requirements referring GitHub repository's name are more intense, while other requirements' distributions are sparser, with an average value around 1120 days. Combining the data analysis in the box plot in Fig.~\ref{fig010}, we observe that the median of response time of requirements referring the GitHub repository's name is around 4 days, therefore about half requirements response time is less than 4 days. 75\% requirements' response time is less than 300 days. Except for a few outliers, this type of requirements response time is within 1000 days. The median of other requirements' response time is around 1000, the first 25\% and the last 25\% of the median are distributed evenly. Only about 25\% of other requirements' response time is less than 300 days. In summary, about 75\% requirements referring repository's name are significantly shorter than other requirements.

Therefore, the requirements referring repository's name has a shorter response time than other requirements. This shows that when discussing the OSS projects in the SO community, if you add a direct reference to GitHub repository, such as repository's name, repository's URL, etc., it will be more helpful for developers to understand and gather the cross-community requirements. 

In addition, in Q\&A community, there is generally a tag system that can classify the questions in more detail and get specific search results, which is convenient for users to quickly browse and locate a certain type of content. Of the 251 posts referring repository's name, 90 of them use repository's name as a tag, which shows that about 1/3 of users have agreed to this behavior and actively used in practice. Therefore, we believe that using the GitHub repository name as a tag will make it easier for developers to search by tag and quickly locate requirements.

\textbf{Citing multiple SO links in a GitHub issue of gathering cross-community requirements: }Another interesting behavior is issue's author citing multiple Q\&A, in order to enrich the understanding of software requirements and help developers comprehend in more detail about the cross-community requirements in GitHub issues. Fig.~\ref{fig011} shows an issue of gathering requirements with multiple links, GitHub developer cites three links in an issue, two of which are from SO. After browsing the issue content, we find that the first SO link is used to gather requirements, and the second SO link provides a possible solution. In order to better understand this behavior, we conduct the following investigation: How many links are cited in cross-community requirements? How many of them are SO links? What is the purpose of citing these SO links? The specific analysis is as follows:

\begin{figure}[htbp]
	\centerline{\includegraphics[width=3.3in]{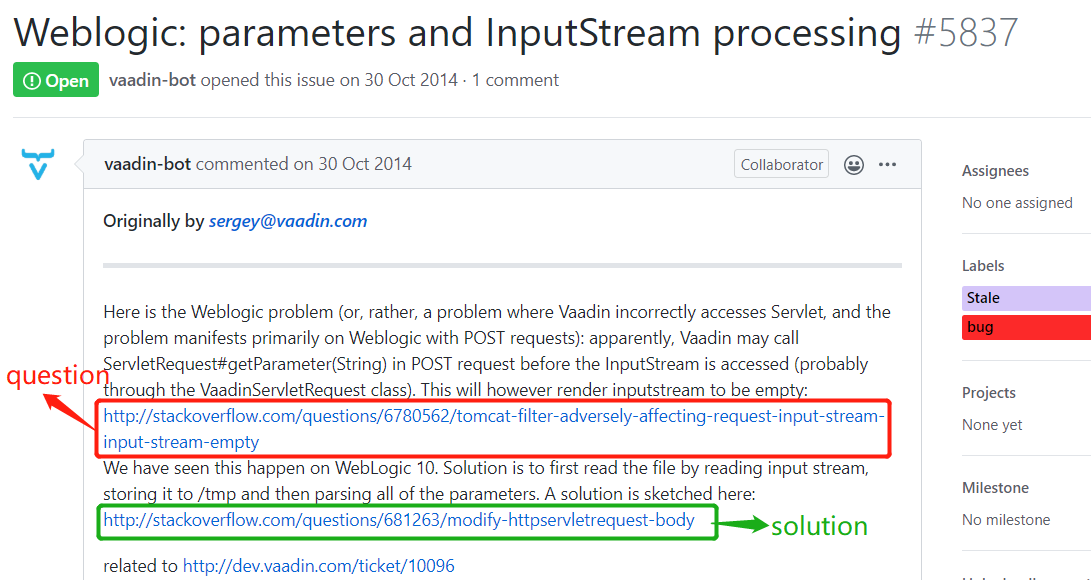}}
	\caption{An example of citing multiple links in cross-community requirements.}
	\label{fig011}
\end{figure}
We analyze the number of links and SO links cited in these issues containing cross-community software requirements. In our research dataset, there are 1873 links and 1334 SO links in 939 issues. As shown in Fig.~\ref{fig012}, there are at most 23 links in the same issue, of which cited up to 18 SO links. We can infer from Fig.~\ref{fig012} that about half (537) of the cross-community requirements contain only one link which is SO link, and the other half contain multiple links, of which the distribution of the number of SO links is shown in Fig.~\ref{fig012}. In addition, we find the behavior with more than 14 links in the same issue is less common.

\begin{figure}[htbp]
	\centerline{\includegraphics[width=3in]{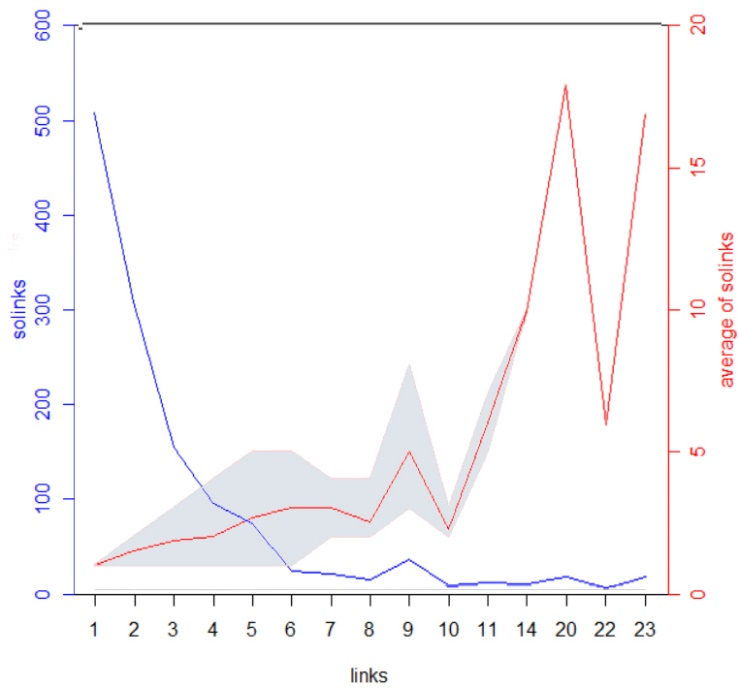}}
	\caption{Analysis of multiple SO links in the same issue.}
	\label{fig012}
\end{figure}
\begin{figure}[htbp]
	\centerline{\includegraphics[width=3in]{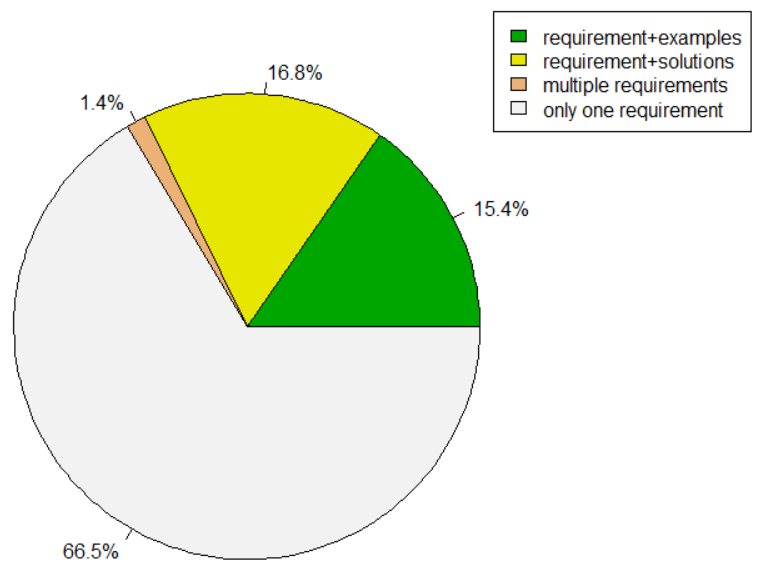}}
	\caption{Proportion of requirements contains multiple links with different purpose.}
	\label{fig013}
\end{figure}

Many developers list multiple requirements, some potential solutions, examples from SO when they create issues, which is also the main reason for including multiple links in an issue. This is one of the benefits of gathering requirements from Q\&A community, which can easily provide many references. To understand the purpose and prevalence of quoting multiple links in an issue, we use regular expressions (solution, workaround, trial, approach, case, example, etc.) to filter the issue content and simply process the data. The results are shown in Fig.~\ref{fig013}, there are 1.4\% issues contains multiple requirements, 15.4\% issues provide examples and 16.8\% issues provide solutions. This is a good cross-community behavior, which can easily provide ideas and clues for OSS developers and help them solve problems quickly.

\noindent
\fbox{%
	\parbox{3.35in}{%
			\textit{Finding 4:} Referring GitHub OSS repository's name in SO posts can shorten the response time of gathering cross-community requirements. Moreover, OSS developers often cite multiple SO posts in GitHub issues, which contain software requirements, solutions, and examples and help developers solve problems quickly.
	}%
}
\section{Discussion}
This section discusses some implications in our research and limitations of our methods and experiments in order to provide experience for future research.
\subsection{Implications for Gathering Requirements}
We discovered that the label of issues of gathering cross-community requirements generically are enhancement, bug, feature, while labels like document, discussion, note, etc. usually do not contain requirements. Furthermore, most requirements gathered from SO are non-functional requirements, such as security issues, usability issues, etc. We notice that these issues are ambiguous when it first proposed in SO posts. Requirements were clearly and directly exposed after fully discussion in SO, and fed back to GitHub repository by users in GitHub. We also detect that there are multi-linking behaviors when gathering requirements from the Q\&A community, which lead to issues with few descriptions but containing large information, and it is helpful for developers to understand the requirements and promote problems solving. Cross-community practice has enabled more people to participate in activities on software development and testing, which has a positive impact on the development of OSS.

\subsection{Threats to validity}
Our study has three main threats.
In our study, we used the manual annotation method to identify cross-community requirements between GitHub and SO, which may not get very good results. Different understanding may cause different results. Moreover, due to the time-consuming of manual annotation, the total size of the samples is too small to accurately capture all software requirements, which poses a threat to the validity of results. 

We extract the title and body of the issues to build the text data, annotating the text to find out the cross-community requirements. However, the data crawled by GitHub API contains issues’ comments citing SO links. Due to the abandonment of this part of the data, we may finally get lower proportion of requirements gathered across communities than the actual situation.

Besides, we observed that some issues contain multiple SO links. We only analyze the first SO link, which may affect the results. The reason for analyzing the first link is that through observation, we find that other links may provide some reference cases or solutions for requirements, not the requirements itself. It may affect the results to a certain extent.

\section{Conclusion}
In this paper, we explore the characteristics and collaborations of gathering GitHub software requirements from SO to proposed some insights behind this phenomenon. We obtained software requirements gathered across GitHub and SO through manual annotation. We analyzed the information of users, links, projects, issues and mines developers' collaboration data in these two communities and makes some valuable findings. 

We find that GitHub’s practice of gathering software requirements from SO started in 2011, grew fastest in 2015, and has achieved stable growth in recent years. Many OSS projects have gathered software requirements across communities. By analyzing the characteristics of requirements gathered from SO, the users are more inclined to gather enhancement requirements, and these requirements are more focused on non-functional requirements.

In addition, we find that trusted users in SO and repository owners in GitHub are more active in gathering requirements from SO. Generally, gathering software requirements across communities requires developers to spend more time, but time also depends on the behavior of participants. Users in SO discuss requirements of OSS projects referring to GitHub repository's name will help developers gather requirements with shorter response time. Moreover, OSS developers in GitHub often cite multiple SO posts, which contain software requirements, solutions, and examples, etc., and play an important role in software development.
In the future, we plan to capture the link information of each other in GitHub and SO, especially identify the requirements of software projects, and make a tool for two-way recommendation.

\section*{Acknowledgment}
This work was supported in part by the National Key Research and Development Program of China under Grant 2018YFB1004202, and in part by the Laboratory of Software Engineering for Complex Systems. 

\vspace{12pt}

\end{document}